\documentclass[aps,prl,twocolumn,superscriptaddress,floatfix]{revtex4-2}
\usepackage{amssymb}
\usepackage{physics}
\usepackage{graphicx}
\usepackage{amsmath}
\usepackage{amsthm}
\usepackage{amsfonts}
\usepackage[T1]{fontenc}
\usepackage{array}
\usepackage{multirow}
\usepackage{color}
\usepackage{esint}
\usepackage{bm}
\usepackage{color}
\usepackage{bbm}
\usepackage{hyperref}
\usepackage{babel}

\newcommand{\EMBL}{``early'' MBL}

\newcommand{\todo}[1]{{\color{cyan} {#1}}}
\newcommand{\lettersection}[1]{\emph{\todo{#1}}.---}

\usepackage{hyperref}
\hypersetup
{	colorlinks,%
	citecolor=green,%
	linkcolor=red,%
	urlcolor=blue%
}
\allowdisplaybreaks[4]
\begin{document}

\global\long\def\id{\mathbbm{1}}
\global\long\def\ui{\mathbbm{i}}
\global\long\def\ud{\mathrm{d}}

\title{Interaction-enhanced many body localization in a generalized Aubry-Andre model}

\author{Ke Huang}
\affiliation{Department of Physics, City University of Hong Kong, Kowloon, Hong Kong SAR, China}

\author{DinhDuy Vu}
\affiliation{Condensed Matter Theory Center and Joint Quantum Institute, University of Maryland, College Park, Maryland 20742, USA}

\author{Sankar Das Sarma}
\affiliation{Condensed Matter Theory Center and Joint Quantum Institute, University of Maryland, College Park, Maryland 20742, USA}

\author{Xiao Li}
\email{xiao.li@cityu.edu.hk}
\affiliation{Department of Physics, City University of Hong Kong, Kowloon, Hong Kong SAR, China}
\date{\today}

\begin{abstract}
We study the many-body localization (MBL) transition in a generalized Aubry-Andre
model (also known as the GPD model) introduced in Phys. Rev. Lett. \textbf{114}, 146601 (2015) (Ref.~\cite{Ganeshan2015}). 
In contrast to MBL in other disordered or quasiperiodic models, interaction seems to unexpectedly enhance MBL in the GPD model in some parameter range. 
To understand this counter-intuitive result, we demonstrate that the highest-energy single-particle band in the GPD model is unstable against even infinitesimal disorder, which leads to this surprising MBL phenomenon in the interacting model.
We develop a mean-field theory description to understand the coupling between extended and localized states, which we validate using extensive exact diagonalization and DMRG-X numerical results.
\end{abstract}

\maketitle

\lettersection{Introduction}
Interacting many-body systems should manifest eigenstate thermalization hypothesis (ETH) with long-time thermalization as the system becomes ergodic~\cite{Deutsch1991,Srednicki1994,Rigol2008}. 
ETH, however, seems to fail generically in disordered or quasiperiodic 1D systems, at least for finite-size systems, although the situation is unclear in the thermodynamic limit. 
Such a phenomenon is known as many-body localization (MBL)~\cite{Nandkishore2015,Altman2015,Abanin2019}. 
MBL has been numerically and experimentally verified in numerous 1D disordered and quasiperiodic interacting systems with the generic finding that the interacting system undergoes an MBL transition at large disorder for a fixed interaction strength. 
In general, the disorder strength necessary for inducing MBL is much larger in the interacting system (and increases with increasing interaction) than in the corresponding noninteracting system.  
For example, the noninteracting 1D Anderson model is localized for any finite disorder whereas the corresponding interacting Anderson model undergoes MBL for large disorder~\cite{Imbrie2016}. 
This is expected as interaction tends to thermalize the system by sharing energy and information among the constituents, leading to ergodicity as postulated in ETH (all noninteracting systems are trivially nonergodic whereas the nonergodicity in MBL is nontrivial violating ETH). 
Another example is the well-known Aubry-Andre (AA) quasiperiodic model which has all states localized or extended for a critical ``disorder'' (i.e., quasiperiodic potential strength) larger or smaller than $1$ for the noninteracting model [using convention in Eq.~\eqref{Eq:Potential} below], whereas the corresponding interacting model exhibits MBL for disorder larger than $1.7$~\cite{Schreiber2015,Kohlert2019,Vu2022,Huang2023}.

However, an important question is whether exceptions to the above scenario can be constructed. 
To address this question, here we describe and analyze a surprising counter-intuitive situation arising in the GPD model~\cite{Ganeshan2015}, where finite interactions may lead to enhanced MBL in the sense that MBL occurs ``early'' with the critical disorder strength for the MBL in the interacting GPD model being lower than that in the noninteracting model. 
This is unexpected as interaction is thought to oppose MBL, and not induce it. 
Since the GPD model has already been studied experimentally~\cite{An2021_PRL}, our predictions are directly verifiable in the laboratory. 
Our work provides important new insights into the competition between localized and extended degrees of freedom in interacting many-body dynamics. 

\lettersection{The GPD model}
We start with the GPD model, introduced in Ref.~\cite{Ganeshan2015}: 
\begin{align}
	H_\text{GPD} = \sum_j\left(tc^\dag_jc_{j+1}+\text{H.c.}\right)+\sum_jV_jn_j,\label{Eq:Model}
\end{align}
where $t$ is the hopping strength, which serves as the energy unit in this work. 
The on-site potential is 
\begin{align}\label{Eq:Potential}
	V_j=\frac{2V\cos(2\pi q j+\phi)}{1-\alpha\cos(2\pi q j+\phi)},
\end{align}
where $q=(3-\sqrt{5})/2$ (equivalent to the golden ratio), $V$ is the potential strength, $\alpha\in (-1,1)$, and $\phi\in(0,2\pi)$ is a random phase.

\begin{figure}[!]
\includegraphics[width=\columnwidth]{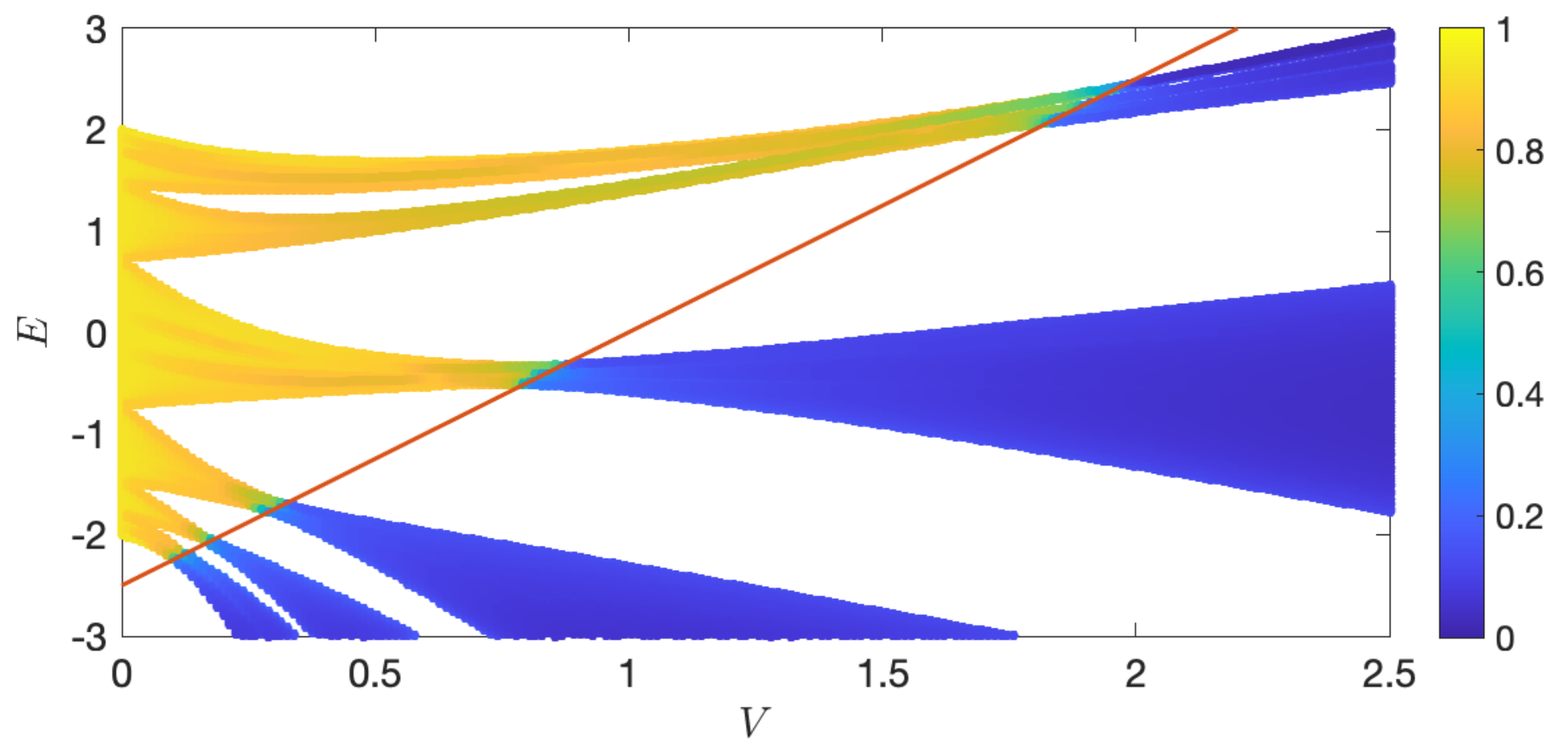}
\caption{\label{Fig:FD}
The fractal dimension of the single-particle GPD model ($U=0$). 
Here, we take $L=610$ and use the periodic boundary condition to avoid localized edge states.
Here (and in all the figures) the $\alpha$ of Eq.~\eqref{Eq:Potential} is $-0.8$, and the line defining SPME is given by $E=5(V-1)/2$. 
}
\end{figure}

For $\alpha=0$, the GPD model reduces to the AA model, which carries no single-particle mobility edge (SPME). 
However, for nonzero $\alpha$, the GPD model is known to carry an exact SPME given by $\alpha E=2(1-V)$~\cite{Ganeshan2015}. 
To show this, we calculate the fractal dimension of the single-particle eigenstates in Fig.~\ref{Fig:FD} for $\alpha=-0.8$. 
The fractal dimension is defined by $\Gamma=-\ln(\sum_j\abs{\psi_j}^4)/\ln L$, where $\psi_j$ is the wave function. 
Consequently, we have $\Gamma\to1$ for extended states and $\Gamma\to0$ for localized states. 
As shown in Fig.~\ref{Fig:FD}, the extended and localized states are separated by the exact SPME, and the system is fully localized at $V_c=2$. 

\lettersection{An {\EMBL}}
We now consider the spinless interacting GPD model at half-filling~\cite{Li2015,Li2016,Deng2017,Hsu2018}. 
Specifically, the interacting GPD Hamiltonian is given by 
$H = H_\text{GPD} + U\sum_jn_jn_{j+1}$, where $U$ is the interaction strength. 
Throughout this work we will take $\phi=0$, $\alpha=-0.8$, and adopt the open boundary condition, unless specified otherwise. 
We use two standard diagnostics to obtain the MBL phase diagram in this model: the entanglement entropy (EE) and the mean gap ratio. 
Here, we use the half-chain second Renyi entropy given by $S=-\ln(\trace_L{\rho_L^2})$ and $\rho_L=\trace_R{\dyad{\psi}}$, where $\trace_L,\trace_R$ denote respectively the partial trace over the left and the right half of the system. 
In addition, the mean gap ratio $\expval{r}$ is defined as the average value of $r_i=\min\{\delta E_i,\delta E_{i+1}\}/\max\{\delta E_i,\delta E_{i+1}\}$, where $\delta E_i=E_{i+1}-E_{i}$ is the energy gap between two adjacent energies. 
In the thermal phase, the EE of the eigenstate approaches the Page value $S_T=(L\ln 2-1)/2$~\cite{Page1993} and $\expval{r}=0.53$ for the Gaussian orthogonal ensemble. 
In the MBL phase we have $S/S_T\to0$ and $\expval{r}=0.38$ for the Poisson distribution. 

\begin{figure}[!]
\includegraphics[width=\columnwidth]{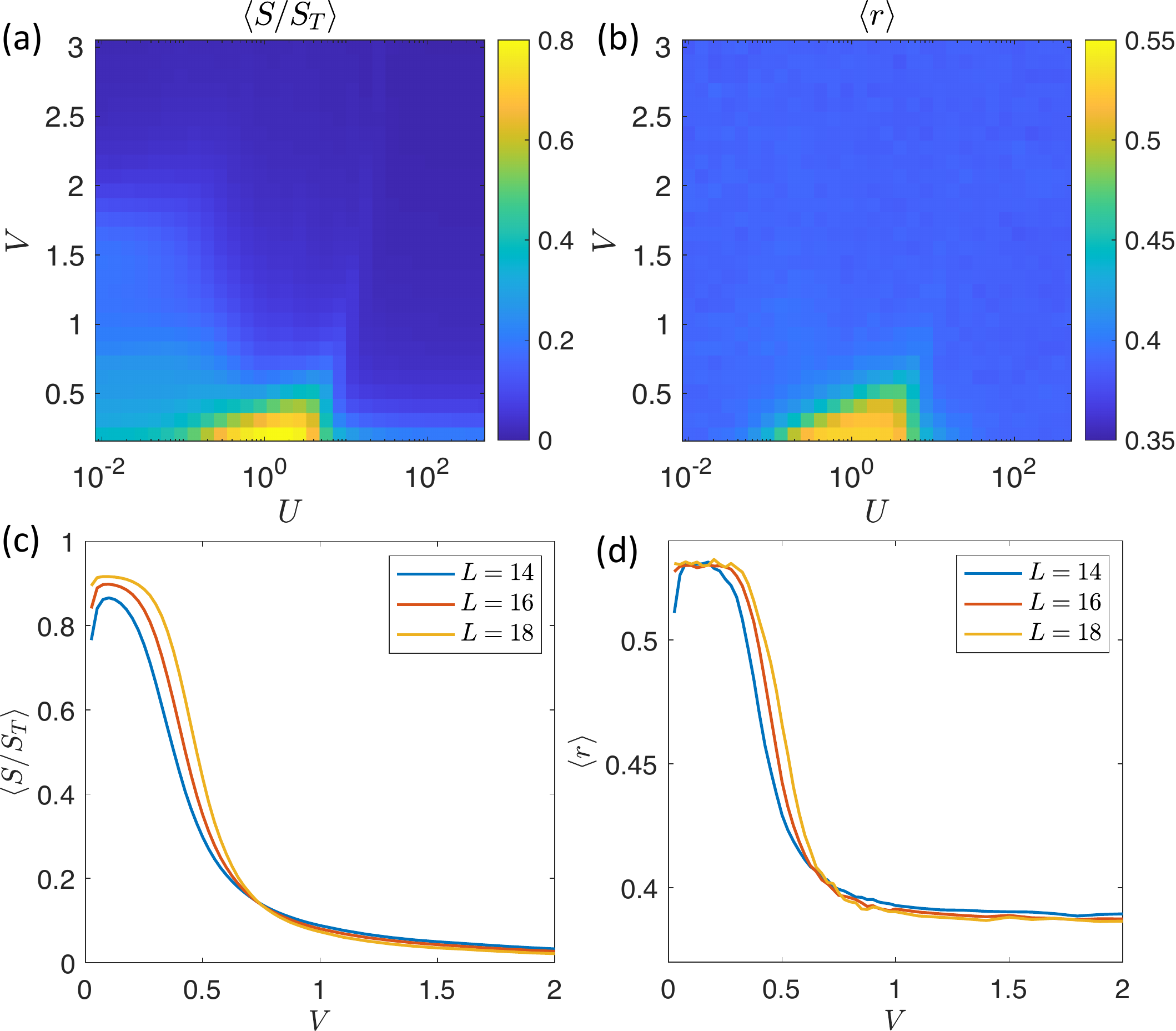}
\caption{\label{Fig:EE_LS} 
(a) and (b) show the entanglement entropy and the mean gap ratio averaged over the whole spectrum in an $L=16$ system as a function of $U$ and $V$. 
(c) and (d) show the same two quantities averaged over the middle quarter of the spectrum for $U=1$ in systems of various sizes. 
The results in (a) and (b) are averaged over $6$ random phase realizations, and the results in (c) and (d) are averaged over $1000$, $200$ and $10$ random phase realizations for $L=14,16$, and $18$, respectively.
}
\end{figure}

In Fig.~\ref{Fig:EE_LS}(a) and~\ref{Fig:EE_LS}(b), we calculate the EE and the mean gap ratio averaged over the whole spectrum in an $L=16$ system. 
In Fig.~\ref{Fig:EE_LS}(c) and~\ref{Fig:EE_LS}(d), we obtain the same quantities for $U=1$ in systems of varying sizes. 
One prominent feature in these results is that the system seems to have an {\EMBL} transition. 
To see this, we focus on the $0.85 < V < 2$ region in Fig.~\ref{Fig:EE_LS}(a) and~\ref{Fig:EE_LS}(b). 
For very weak interactions ($U\ll 1$), the mean gap ratio is always $0.38$, while the EE scales as $S\propto \ln L$ in the presence of extended states. 
Such features are consistent with the single-particle properties of the GPD model. 
However, as the interaction $U$ increases, the system becomes \emph{localized} rather than thermalized, as shown by the drastic decease of the EE. 
This interaction-induced localization is also confirmed by the mean gap ratio. 
To further verify this {\EMBL} transition, we take $U=1$ and analyze the finite-size effect in Fig.~\ref{Fig:EE_LS}(c) and~\ref{Fig:EE_LS}(d). 
Both of them indicate an MBL transition around $V=0.75$, which is much smaller than the full single-particle localization transition at $V_c=2$.
We thus see that the MBL transition in this model happens consistently at a critical disorder strength well below the single-particle localization point.  
Such a feature is in stark contrast with the MBL transition found in all other models, where MBL happens at a critical disorder substantially larger than that for the noninteracting case~\cite{Nandkishore2015,Altman2015,Abanin2019,Imbrie2016,Vu2022,Huang2023}.

\lettersection{Fragile flat bands}
To explain this surprising {\EMBL} transition, we take a closer look at the single-particle properties of the GPD model. 
From Fig.~\ref{Fig:FD} we see that for $V\in(0.85,2)$ most eigenstates in the model are already localized, and only the highest-energy band remains extended. 
Moreover, we find numerically that this flatband contains approximately $qL$ states, so nearly $40\%$ of the states are extended for $V\in(0.85,2)$. 
Moreover, compared to other bands, the width of this extended band is very small. 
It turns out that this extended band is very fragile against external perturbations, which distinguishes the GPD model from some other extensively studied models, such as the AA model~\cite{Supplement}.  

\begin{figure}[!]
\includegraphics[width=\columnwidth]{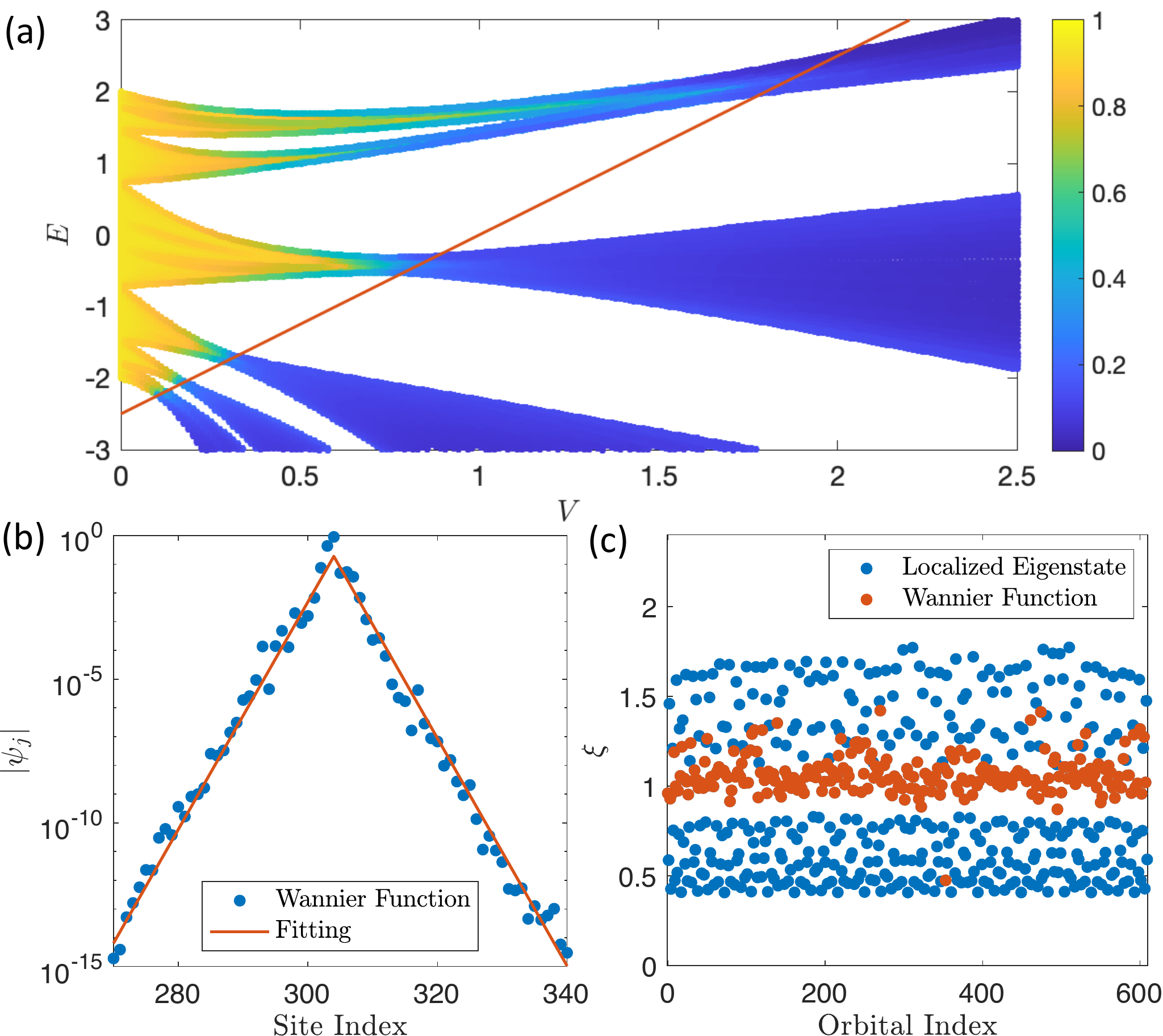}
\caption{\label{Fig:Wannier} 
(a) The fractal dimension of all eigenstates in a single-particle GPD model under the perturbation of Eq.~\eqref{Eq:Disorder}.   
(b) The wave function of one Wannier function and its fitting. 
(c) The localization length of the localized eigenstates and the Wannier functions. Here, the system size is $L=610$. 
}
\end{figure}

We demonstrate this feature by adding to the single-particle GPD model an additional weak random disorder,
\begin{align}
	\delta V=V\sum_{j}\delta h_jn_j,\label{Eq:Disorder}
\end{align}
where $\delta h_j$ is uniformly distributed in $[-1/20,1/20]$. 
We find that, as a result of the above perturbation, the highest-energy band in the GPD model becomes completely localized, causing the full localization transition point to decrease from $V_c=2$ to below $V=1$, as shown in Fig.~\ref{Fig:Wannier}(a). 
Intuitively, this observation can be understood by noticing that the width of the flatband $w$ is much smaller than the band gap $\Delta$ between this flatband and the other states, making this flatband susceptible to perturbations. 
In particular, as long as the disorder satisfies $w\ll\delta V\ll \Delta$, first-order perturbation theory leads to the following effective Hamiltonian,\begin{align}
	H_{\text{eff}}=P\delta VP,
\end{align}
where $P$ is the projection operator of the flatband. 
Therefore, the system tends to localize the flatband states in the presence of any disorder, as shown in Fig.~\ref{Fig:Wannier}(b). 
These localized eigenstates constitute a localized basis of the flatband, so they can be regarded as the Wannier functions for this flatband. 
Keep in mind, however, that there is no standard definition of Wannier function in a quasiperiodic system.
Consequently, the shape of the Wannier functions is highly dependent on the perturbation we select. 
Nonetheless, the localization centers of any two Wannier functions are separated by at least two lattice sites, regardless of the choice of perturbation.  
To quantify the extent of localization, we extract the localization length $\xi$ of the Wannier functions by fitting the wave function to $\psi_j\propto \exp(-\abs{j-j_0}/\xi)$, as shown in Fig.~\ref{Fig:Wannier}(b). 
Our results in Fig.~\ref{Fig:Wannier}(c) show that all the Wannier functions are deeply localized states with $\xi\sim1$.
Hence, the single-particle GPD model with $\alpha = -0.8$ resembles a deeply localized system, disguised by weak tunnelling that can be destroyed by disorder. 
In fact, the fragile flatband is only prominent for large $\alpha$ (i.e., $\abs{\alpha}$ close to 1) and small $q$, and therefore our discussion below is strictly carried out in this limit.

\lettersection{A mean-field description for the transition}
The {\EMBL} transition suggests that the extended orbitals in the flatband, amounting to $40\%$ of all the orbitals, are localized by the other localized single-particle states.  
Based on our discovery of the fragile flatband in the GPD model, an intuitive argument for the {\EMBL} transition is that quantum fluctuations coming from the localized orbitals serve as the additional disorder and localize the extended flatband as the interaction is turned on. 
This heuristic viewpoint can be theoretically formulated by a mean-field (MF) theory. 
The MF theory is essentially a set of nonlinear self-consistent equations, and the physics can be explained by the MF theory if the solution of the self-consistent equations agrees with the numerically calculated many-body eigenstates. 

Let us construct a complete and deep-localized basis utilizing the Wannier functions together with the other localized orbitals. 
Each lattice site can be associated with a unique basis orbital localized on this site. 
Under this basis, the model of Eq.~\eqref{Eq:Model} can be rewritten as
\begin{align}
	H_\text{GPD}^\text{(new)}=&\underbrace{\sum_{j}\epsilon_j\tilde n_j+\sum_{i\neq j}U_{ij}\tilde n_i\tilde n_j}_{\text{dominant}}\nonumber\\
	&+\underbrace{\sum_{\substack{i,j\in \text{flat}\\i\neq j}}t_{ij}f^\dag_if_j+\sideset{}{'}\sum_{ijkl}U_{ijkl}f^\dag_if^\dag_jf_kf_l}_{\text{perturbative}},\label{Eq:Perturbation}
\end{align}
where $f_k$ annihilates the basis orbital localized on site $j$, and $\tilde n_j=f^\dag_jf_j$ is the particle number of the orbital $j$. 
In addition, $j\in\text{flat}$ denotes the Wannier functions of the flatband, and $\sideset{}{'}\sum_{ijkl}$ denotes that three of the four indices $ijkl$ are different. 
The first term in Eq.~\eqref{Eq:Perturbation} is the energy of the orbital, and the second is the diagonal part of the interaction in this basis, serving as the additional disorder. 
These two terms contribute the dominant part of the Hamiltonian. 
The third term comes from the weak tunnelling between the Wannier functions, and we estimate that $t_{ij}\sim w/2\sim O(10^{-1})$. 
Finally, the last term is the off-diagonal part of the interaction, and thus
\begin{align}
\begin{aligned}
U_{ijkl} & \sim U\exp[-(\max\{i, j, k, l\}-\min\{i, j, k, l\})/\xi] \\
&\lesssim U\exp(-2/\xi)\sim \order{10^{-1}}. 
\end{aligned}
\end{align}
Hence, the last two off-diagonal terms are perturbative, suggesting that the MF theory should work well. 
Although we choose a specific basis in this argument, the self-consistent equations are basis-independent, and so are the MF solutions. 
Our numerical results below show that the precision of the MF solution is much better than $\order{10^{-1}}$. 
The above argument also applies to MBL in other models, and we specifically show for the AA model~\cite{Supplement}. 

\lettersection{Accuracy of the MF solution}
We now demonstrate numerically the accuracy of the MF theory.  
We fix $V=1.5$, which is smaller than the single-particle localization transition point.  
The standard MF theory is targeting the low-temperature physics, and thus we need to calculate the ground state of the self-consistent equations. 
However, to analyze the {\EMBL} transition, we must obtain the highly excited solution of the self-consistent equations, which is notoriously difficult in generic models. 
In the present problem, when the system is in the MBL phase, we are able to construct an efficient algorithm to solve the MF equations for the highly excited states, as we now explain. 
To obtain the ground state, one generally starts from an initial product state, replaces the nonlinear terms with the solution in the previous round of iteration, and minimizes the energy in each round of iteration. 
To derive the highly excited states, in contrast, we start from the product state of the localized orbitals, replace the nonlinear terms, and maximize the overlap between the next state and the current state, a procedure resembling the DMRG-X algorithm~\cite{Khemani2016,Devakul2017}. 
If the iteration converges, we then generate an MF state from the initial Fock state. 
We emphasize that despite the similarity between DMRG-X and our modified MF theory, these two algorithms focus on very different aspects. 
First, although DMRG-X is expected to provide a more quantitatively accurate approximation, the MF theory provides a more intuitive understanding of the physics behind the numerical results. 
Second, the MF theory is much faster numerically than the DMRG-X algorithm. 
Finally, the MF theory can actually provide more accurate initial states for the DMRG-X algorithm. 

\begin{figure}[!]
\includegraphics[width=\columnwidth]{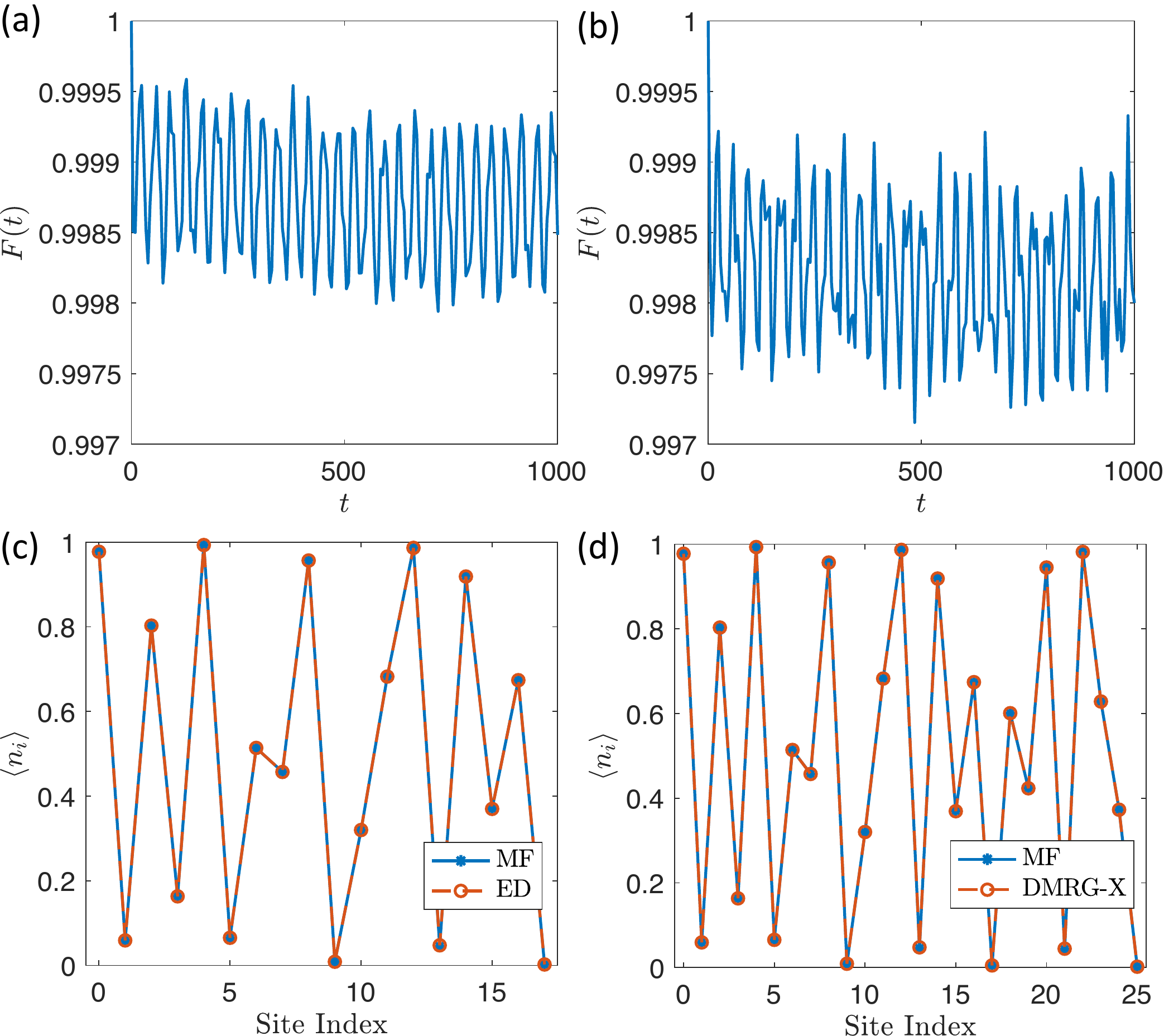}
\caption{\label{Fig:Fidelity}
(a) and (b) show the fidelity $F(t)=\abs{\braket{\psi}{\psi(t)}}$ of the MF N\'{e}el state in an $L=18$ (using ED) and an $L=26$ system (using KPM), respectively. 
(c) and (d) show the expectation of the particle number in the MF N\'{e}el states and their corresponding eigenstates derived by ED or DMRG-X. 
The system size is $L=18$ in (c) and $L=26$ in (d). 
For the DMRG-X algorithm, the energy uncertainty is $\expval{H^2}-\expval{H}^2=2.9\times10^{-9}$. 
Here, we take $U=1$.
}
\end{figure}

To illustrate the power our MF theory, we first study a particular MF solution, the MF N\'{e}el state $\ket{Z_2}$ generated by the N\'{e}el-like Fock state $f_0^\dag f_2^\dag f_4^\dag\cdots\ket{0}$. 
In Fig.~\ref{Fig:Fidelity}, we compute the fidelity of the MF N\'{e}el state $F(t)=\abs{\braket{Z_2}{Z_2(t)}}$ in an $L=18$ system using exact diagonalization (ED) and $L=26$ system using the kernel polynomial method (KPM)~\cite{Weisse2006}. 
The fidelity in both systems remains surprisingly high ($>0.997$) and does not decay after $1000$ tunnelling times. 
This implies that the MF N\'{e}el state is fairly close to one of the many-body eigenstates. 
To compare the MF N\'{e}el state with the eigenstates, we use ED to obtain all eigenstates in the $L=18$ system and choose the eigenstate with the highest overlap  ($=0.9994$) with the MF solution.  
As shown in Fig.~\ref{Fig:Fidelity}, the density profile of the exact eigenstate and that of the MF N\'{e}el state are almost indistinguishable. 
For the $L=26$ system, we utilize the DMRG-X algorithm with the MF N\'{e}el state being the initial state to generate the eigenstate. 
The DMRG-X algorithm gives us a rather accurate result with the energy uncertainty of $\expval{H^2}-\expval{H}^2=2.9\times10^{-9}$. 
Similar to the result in the smaller system, one can barely differentiate the MF and the DMRG-X results. 
We also check that the overlap between the DMRG-X and the MF results is $\abs{\innerproduct{\psi_\text{MF}}{\psi_\text{DMRG-X}}} = 0.9992$. 
Note that in general the density profile in the $L=18$ chain is quite similar to that in the $L=26$ chain, which shows that the system is deep in the MBL phase.

\begin{figure}[!]
\includegraphics[width=0.9\columnwidth]{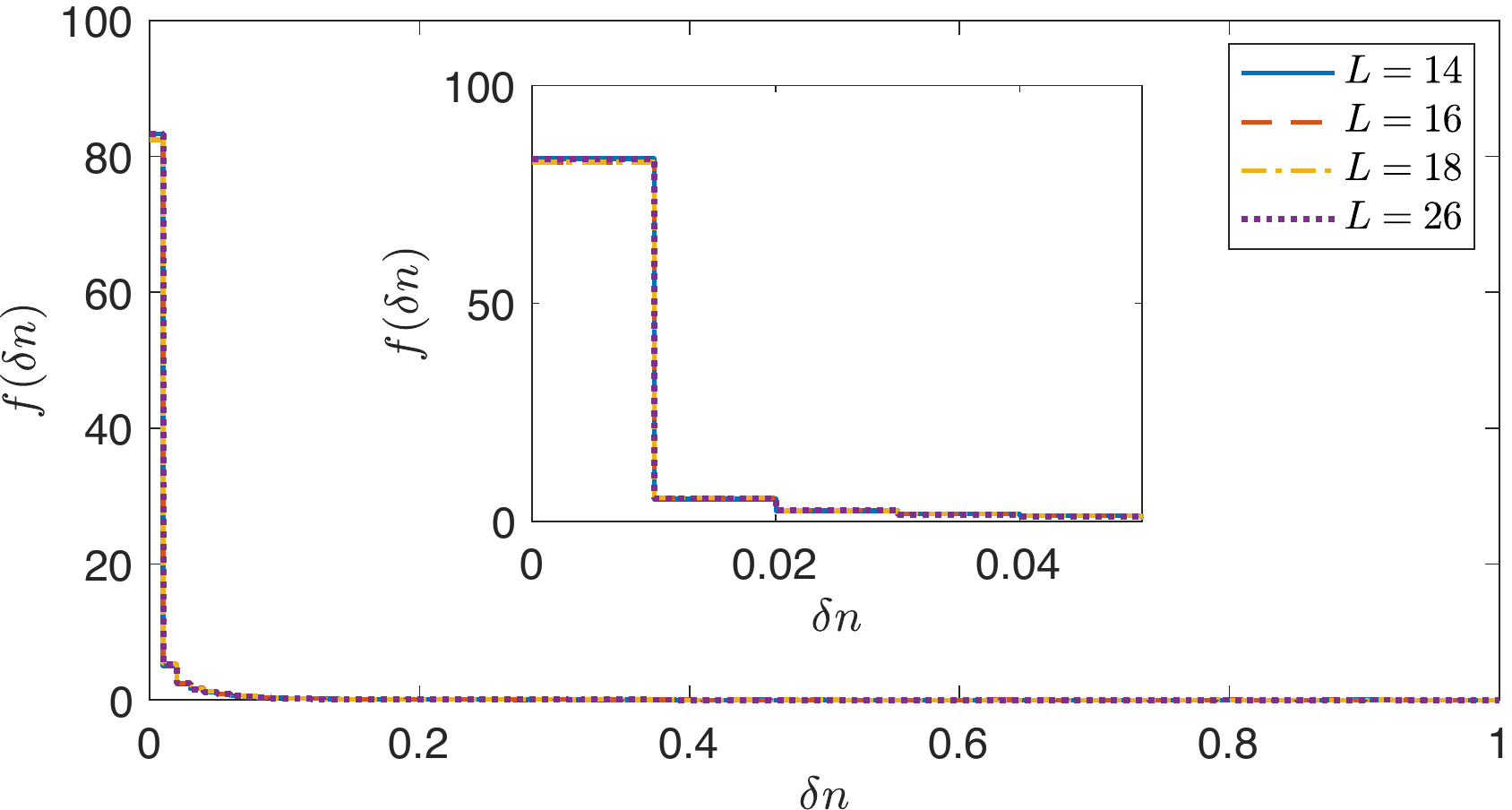}
\caption{\label{Fig:comparison} 
The probability density $f(\delta n)$ of $\delta n$ in the GPD model, and the inset zooms into $\delta n\in[0,0.05]$.  For $L=14,16$ and $18$, we consider all the sites and all the eigenstates obtained by ED. For $L=26$, we consider all the sites and use DMRG-X to generate $2000$ random eigenstates. Here, we take $U=1$.
}
\end{figure}

\lettersection{MF theory for generic states}
To further validate the MF theory, we now investigate the agreement between generic eigenstates and their corresponding MF solutions. 
Given that we are focusing on the MBL regime, it is more appropriate to characterize the quality of the approximation using local quantities, such as particle density, rather than global quantities, such as the overlap. 
Particularly, we plot $f(\delta n)$, the probability density of $\delta n$ in Fig.~\ref{Fig:comparison} with normalization $\int f(\delta) \dd{\delta n} = 1$. 
This quantity is defined as the absolute value of the particle number difference on a random site between a random eigenstate and its corresponding MF state. 
To efficiently obtain the corresponding MF solutions of a given eigenstate, we use the one-body reduced density matrix of the eigenstate $\mel{E}{c_ic^\dag_j}{E}$ as the initial state for the iteration of the self-consistent equations. 
In practice, we find that about $2\%$ of the states do not converge at all, and about $10\%$ of them converge to the MF solution corresponding to the other eigenstates. 
Even including these $12\%$ of solutions, we find in Fig.~\ref{Fig:comparison} that $\delta n$ is highly concentrated around zero. 
What is more, the statistics of the quantity manifests almost no finite-size effects, which can be seen by comparison with the ED results in the smaller systems and the DMRG-X results in the larger system. 
Hence, the numerical results verify that the MF theory is a general explanation for the {\EMBL} transition in the GPD model.

\lettersection{Conclusion}
In this work we studied a surprising interaction enhanced MBL, which is a special feature of the GPD model. 
This phenomenon is explained by a MF theory construction, which is validated by comparison with numerical results obtained by ED and DMRG-X algorithms. 
Given the recent experimental implementation of the GPD model~\cite{An2021_PRL}, our predictions can be experimentally verified.

\lettersection{Acknowledgement}
This work is supported by the Laboratory for Physical Sciences, the Research Grants Council of Hong Kong (Grants~No.~CityU~21304720, CityU~11300421, and C7012-21G), and City University of Hong Kong (Project~No.~9610428).  
K.H. is also supported by the Hong Kong PhD Fellowship Scheme.

\bibliographystyle{apsrev4-2}
\bibliography{EMBL_v5b.bib}

\end{document}


\global\long\def\id{\mathbbm{1}}
\global\long\def\ui{\mathbbm{i}}
\global\long\def\ud{\mathrm{d}}

\title{Supplemental Material for ``Interaction-enhanced many body localization in a generalized Aubry-Andre model''}

\author{Ke Huang}
\affiliation{Department of Physics, City University of Hong Kong, Kowloon, Hong Kong SAR}
\author{DinhDuy Vu}
\affiliation{Condensed Matter Theory Center and Joint Quantum Institute, University of Maryland, College Park, Maryland 20742, USA}
\author{Sankar Das Sarma}
\affiliation{Condensed Matter Theory Center and Joint Quantum Institute, University of Maryland, College Park, Maryland 20742, USA}
\author{Xiao Li}
\email{xiao.li@cityu.edu.hk}
\affiliation{Department of Physics, City University of Hong Kong, Kowloon, Hong Kong SAR}

\setcounter{equation}{0} 
\setcounter{figure}{0}
\setcounter{table}{0} 
\renewcommand{\theparagraph}{\bf}
\renewcommand{\thefigure}{S\arabic{figure}}
\renewcommand{\theequation}{S\arabic{equation}}
\renewcommand\labelenumi{(\theenumi)}

\maketitle

\section{Stability of the band of extended states in the GPD model\label{Appendix:Instability}}

\begin{figure}[b]
\includegraphics[width=\columnwidth]{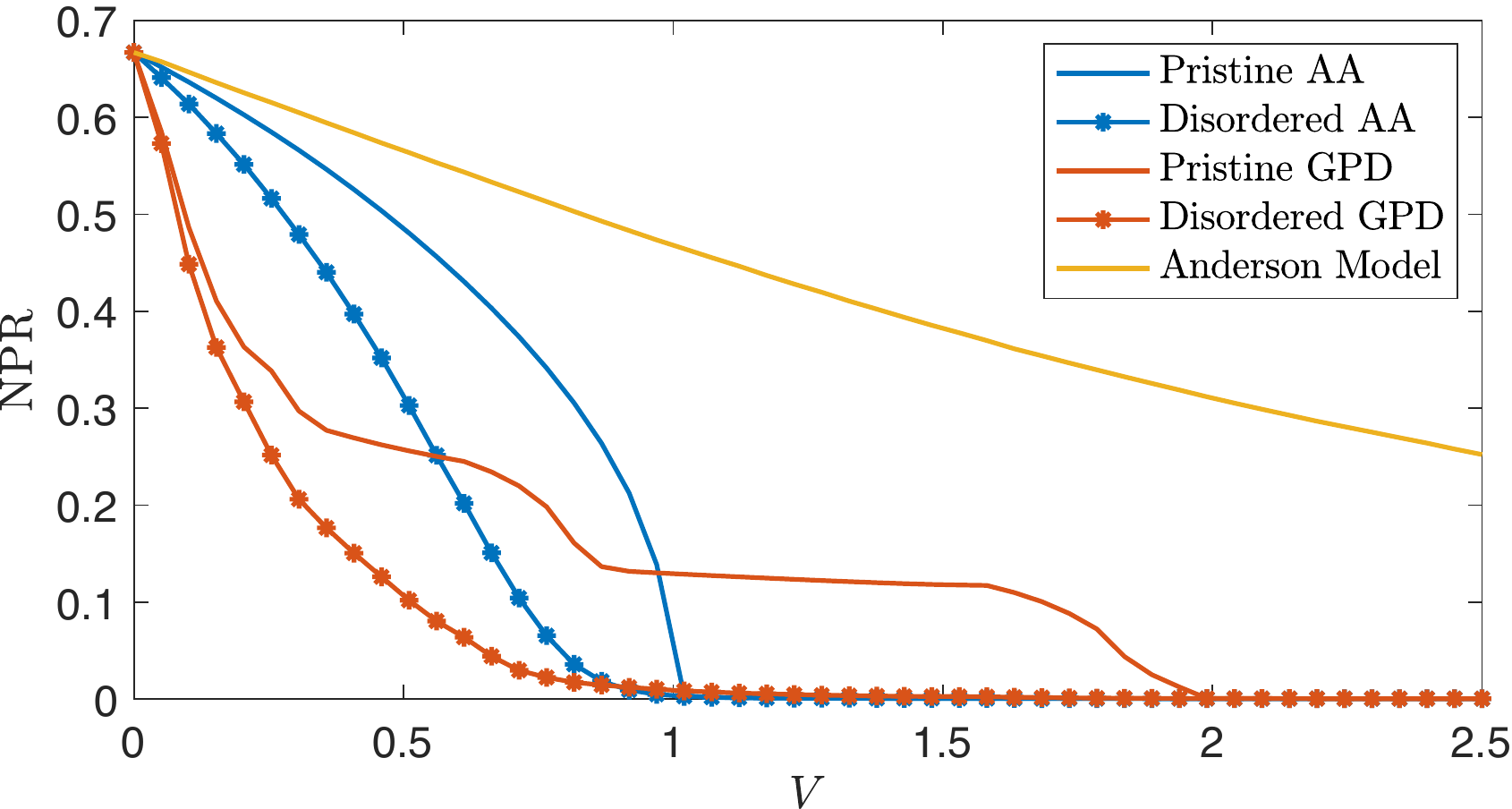}
\caption{\label{Fig:NPR} 
NPR of single-particle eigenstates in the AA model, the GPD model, and the Anderson model. 
Here, the system size is $L=2584$. 
}
\end{figure}

We now show that the existence of an extended and narrow band of eigenstates in the GPD model is an intrinsic and unusual property of this model. 
We illustrate this point by comparing the GPD model with the AA model and the Anderson model. 

The pristine model considered here is 
\begin{align}
	H=&\sum_j\left(tc^\dag_jc_{j+1}+\text{H.c.}\right)+\sum_jV_jn_j,
\end{align}
where $V_j$ is given in the main text. We still take $\alpha=-0.8$ for the clean GPD model, while the clean AA model is the $\alpha=0$ case. Their corresponding disordered models are obtained by adding $\delta V$ introduced in in the main text. Further, we also study the weak disordered Anderson model as a benchmark,
\begin{align}
	H=\sum_j\left(tc^\dag_jc_{j+1}+\text{H.c.}\right)+\delta V.
\end{align}
The instability of the extended states can be signified by an earlier full localization in the disorder model. To identify full localization, we introduce the normalized participation ratio (NPR), defined by
$\text{NPR}=1/\left[L\sum_j\abs{\psi_j}^4\right]$
where $\psi_j$ is the wave function. The NPR vanishes for localized states, and therefore the NPR averaged over eigenstates becomes zero only if the system is fully localized. In Fig.~\ref{Fig:NPR}, we calculate the averaged NPR for the AA model, the GPD model, and the Anderson model. The full localization transition is significantly advanced in the GPD model, whereas the weak disorder can hardly change the transition in the AA model. Although infinitesimal disorder can localized 1D systems, weak disorder will result in a very long localization length, and the system still seems extended in a finite small system as shown in Fig.~\ref{Fig:NPR}. This is in clear contrast to the disorder GPD model where the system is obviously localized even in a small system. Hence, the instability should be an intrinsic property of quasiperiodic systems.

\section{MF theory in the AA model}\label{A:AA}

\begin{figure}[t]
\includegraphics[width=\columnwidth]{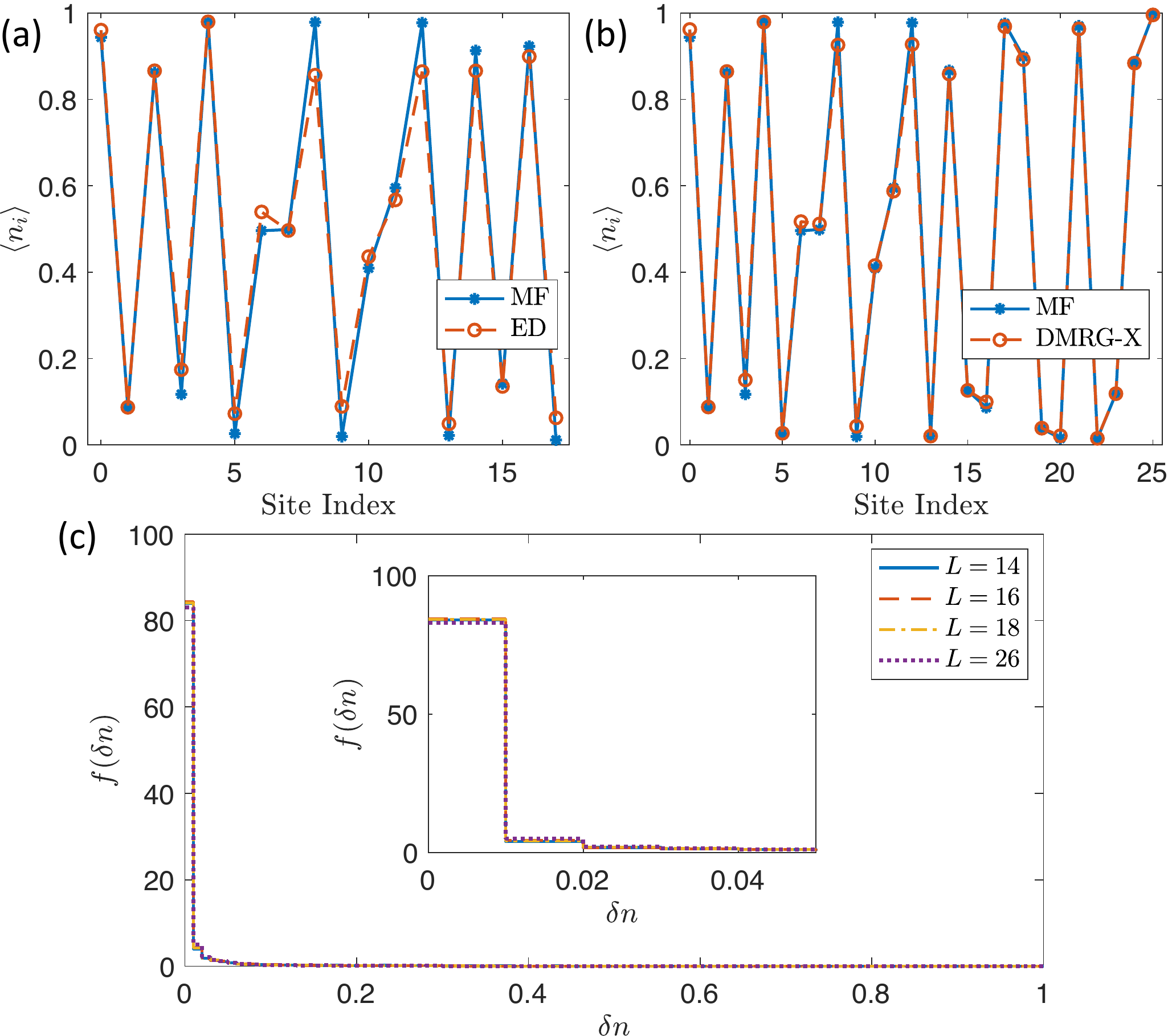}
\caption{\label{FigSM:comparison} 
(a) and (b) show the expectation of the particle number in the MF N\'{e}el states and their corresponding eigenstates derived by ED or DMRG-X. 
The system size is $L=18$ in (a) and $L=26$ in (b). 
For the DMRG-X algorithm, the energy uncertainty is $\expval{H^2}-\expval{H}^2=3.8\times10^{-7}$. 
(c) The probability density $f(\delta n)$ of $\delta n$ in the AA model, and the inset zooms into $\delta n\in[0,0.05]$.  For $L=14,16$ and $18$, we consider all the sites and all the eigenstates obtained by ED. For $L=26$, we consider all the sites and use DMRG-X to generate $2000$ random eigenstates. Here, we take $U=1$.
}
\end{figure}

The MF argument in the main text is based on the existence of a set of deep-localized orbitals and perturbative off-diagonal terms, which is generally satisfied in the deep MBL phase of various models. 
To test the universality of the MF theory, we study the MBL phase in the AA model. 
The MF theory in the AA model is much simpler than that in the GPD model. 
First, as the AA model has no mobility edge, it is impossible that the localized states serve as disorder. 
Second, even if there are external random disorders, the AA model is stable against weak random disorders as shown in Fig.~\ref{Fig:NPR}. 
Hence, the MBL transition in the AA model is much later than the single particle transition, meaning that all single particle orbitals are deep localized states and that we do not need to construct Wannier functions.

Here, we take $V=2.5$ because the AA model with $U=1$ is known to be localized around $V\approx1.7$. In Fig.~\ref{FigSM:comparison}(a) and~\ref{FigSM:comparison}(b), we calculate the MF N\'{e}el state in the AA model, and compare them with the ED and the DMRG-X. The high overlap ($0.9260$ for $L=18$, and $0.9644$ for $L=26$) and the agreement of the density profile validate the MF argument in the AA model. Note that similar to the GPD model in the main text, the particle density on the first 16 sites in the $L=26$ system resembles that in the $L=18$ system. This is a direct consequence of the localization because the particles on the left can hardly feel the effect of the additional patch from site $17$ to site $26$ on the right. Nonetheless, the region close to site $17$ can still feel the additional patch, so the density close to site $17$ is general different as shown by Fig.~\ref{FigSM:comparison}(a) and~\ref{FigSM:comparison}(b). In fact, it is a coincidence that the MF N\'{e}el states in the GPD model even agrees well around site $17$. Furthermore, we also calculate the $\delta n$ in Fig.~\ref{FigSM:comparison}(c). The result is quantitatively similar to Fig.~\ref{Fig:comparison}, manifesting an excellent precision of the MF solutions.

\section{The ineffectiveness of the NN interaction}

\begin{figure}[b]
\includegraphics[width=\columnwidth]{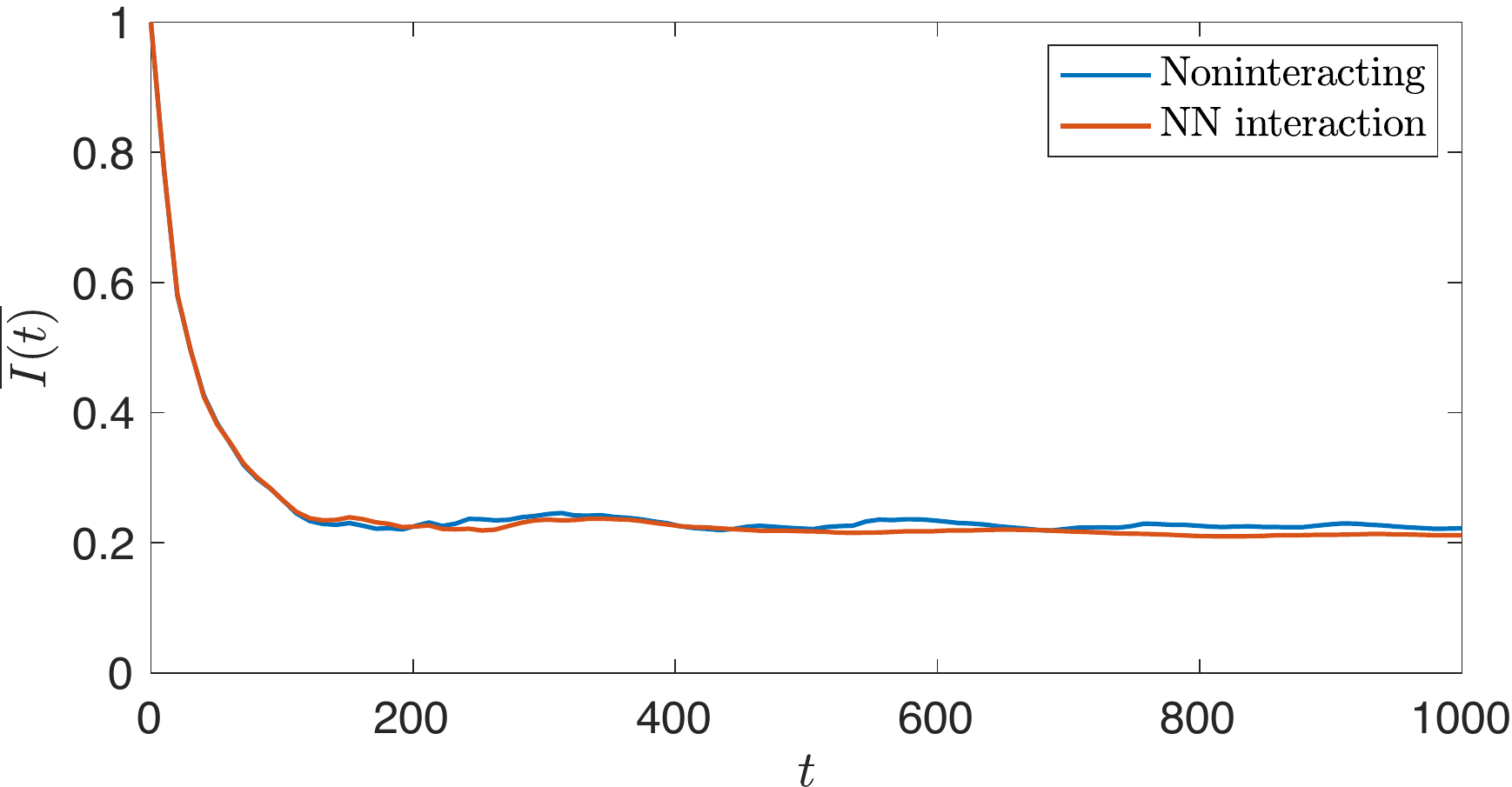}
\caption{\label{FigSM:Dyn} 
Time-averaged imbalance for the noninteracting and the NN interaction cases. Here, we take $L=21$ and $N=4$ for half-filled flatband, and $U=1$ for the NN interaction.
}
\end{figure}

In the main text, we demonstrate the localization effect caused by the localized single-particle orbitals. However, if the state only occupies the extended orbitals, then there is no effective disorder to localize the system. Note that this phenomenon never appears at half filling, because the flatband only contains $38\%$ of the orbitals and some localized orbitals must be occupied. Moreover, as the localization centers of the Wannier functions are at least separated from each other by two sites, the NN interaction has very weak effect within the flatband. Hence, we anticipate that the state remains extend in the absence of the localized orbitals. 

To numerically confirm it, we study the system of length $L=21$, which possesses $8$ extended orbitals in the flatband, and calculate the following imbalance dynamics,
\begin{align}
	I(t)=N^{-1}\sum_{i\in\text{flat}}\trace[P_{\text{flat}}\tilde n_i(t)\tilde n_i],
\end{align}
where $\tilde n_i(t)$ is the particle number operator $\tilde n_i$ in the Heisenberg picture. 
In Fig.~\ref{FigSM:Dyn}, we calculate the time-averaged imbalance $\overline{I(t)}=t^{-1}\int_0^tI(s)\dd s$. Both the noninteracting case and the NN interaction case relax from $I=1$ to $I=0.2$ at late time. Moreover, they manifest similar dynamics, validating the weak effect of the NN interaction within the flatband.

\bibliographystyle{apsrev4-2}
\bibliography{t1t2_MBL_v2}